\documentclass[eqsecnum,twocolumn,showpacs,amsmath,amssymb]{revtex4}
\usepackage{bm}
\usepackage{graphicx}
\usepackage{amsfonts}

\newcommand{\gd}{\dot\gamma}
\newcommand{\bx}{\boldsymbol x}
\newcommand{\bq}{\boldsymbol q}
\newcommand{\bqp}{{\boldsymbol q_{\perp}}}
\newcommand{\bk}{{\boldsymbol k}}

\newcommand{\Om}{\Omega}
\newcommand{\om}{\omega}
\begin{document}
\title{Dynamical Coarse-Graining of Highly Fluctuating Membranes under
  Shear Flow} 
\author{Simon W. Marlow}
\email{physwm@irc.leeds.ac.uk} 
\author{Peter D. Olmsted}
\email{p.d.olmsted@leeds.ac.uk}
\affiliation{Polymer IRC and
  Department of Physics \& Astronomy, University of Leeds, Leeds LS2
  9JT, United Kingdom} 
\date{\today} 
\begin{abstract}
  The effect of strong shear flow on highly fluctuating lamellar
  systems stabilized by intermembrane collisions via the Helfrich
  interaction is studied. Advection enters the microscopic equation of
  motion for a single membrane via a non-linear coupling.  Upon
  coarse-graining the theory for a single bilayer up to the length
  scale of the collision length, at which a hydrodynamic description
  applies, an additional dynamical coupling is generated which is of
  the form of a wavevector-dependent tension that is non-linear in the
  applied shear rate.  This new term has consequences for the effects
  of strong flow on the stability and dynamics of lamellar surfactant
  phases.
\end{abstract}
\maketitle
\section{Introduction and Overview}
Dilute solutions of lamellar phases typically consist of highly
fluctuating layers. The wide equilibrium layer spacings are governed
by the interplay between the long-range steric repulsion, known as the
Helfrich interaction \cite{helfrich78}, and the bending elasticity of
the bilayers. When these systems undergo flow, a range of interesting
phenomena is observed, transitions to multilamellar vesicles
\cite{DRN93,SierRoux97} and a reduction in layer spacing
\cite{YamaTana95b}. Unlike layered one component melts, such as
thermotropic smectics or diblock copolymers, flow can have a
significant effect on the microstructure of the layers. Although flow
certainly stretches the chains in diblock copolymers
\cite{williamsmack94} and can induce layer tilt in thermotropic
smectics \cite{AuernhammerBP00}, the effect on the highly-fluctuating
many-component layered surfactant phases should be much more dramatic.

\begin{figure}[htbp]
     \begin{center}
       \includegraphics[width=8cm]{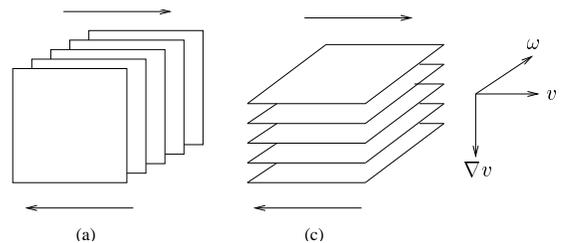}
     \caption{The allowable steady state orientations $a$ and $c$ of a lamellar
       phase in uniform shear flow.}
      \label{fig:0b}
     \end{center}  
\end{figure}

As an initial step to account for some of this flow behaviour, we
previously conjectured \cite{marlowunpub} that flow induces an
effective anisotropic tension parallel to the flow, when lamellae are
aligned in the $c$ orientation (Fig.~\ref{fig:0b}). The ``tension'' is
a response to projected area changes and acts to suppress the
fluctuations. This led to predictions for either changes in layer
spacing or an undulation instability.  In a related work, Zilman and
Granek \cite{zilman99} also proposed an effective tension, but
isotropic, negative in sign, and of a different physical origin. 
While both studies relied on inserting the ``tension'' 
heuristically into the
dynamics as an effective free energy term, it is of interest to
examine the dynamics of a membrane in flow to see how such a response
can be generated dynamically. In this work we consider a lamellar
phase stabilized by the Helfrich interaction, neglecting the effect of
electrostatic forces. By coarse-graining the
dynamics up to a length scale characteristic of the long wavelength
hydrodynamic description, the typical transverse length $L_p$ between
collisions, we demonstrate that flow can indeed induce a dynamical
suppression of fluctuations that resembles a wavevector-dependent
``tension''.

The linearized relaxational dynamics of a membrane parametrized
by a single height variable $h$ subject to an
anisotropic tension $\sigma$, written in Fourier space, is
\begin{equation}
  \label{eq:0b}
  \partial_t h(\bq) = 
  -\Lambda(\bq)\left[\kappa q^4 + \sigma q_x^2 \right]h(\bq),
\end{equation}
where $\kappa$ is the bending modulus, $\bq$ is the wave vector 
and the kinetic coefficient
$\Lambda(\bq)$ depends on the particular relaxation mechanism.  The
tension penalizes fluctuations and, if applied to a stack of such
membranes that interact via collisions, would change the ``preferred''
layer spacing and renormalize the coarse-grained smectic layer
compression modulus $\bar B$. Our task here is to derive an equivalent
term that would contribute to the effective dynamics of highly
fluctuating membranes in shear flow in the $c$ orientation, 
${\boldsymbol v}=\gd z\hat{\bx}$. We shall find that
the equation of motion for the coarse-grained height field 
in shear flow becomes
\begin{widetext}  
\begin{equation}
  \label{eq:0c}
  \partial_t h(\bq) + i \dot\gamma \sum_{\bk}(q_x -k_x ) h(\bk) 
  h(\bq - \bk ) =  
  - \left\{\Lambda(\bq )\kappa q^4 + \Lambda_x  q_x^2 \right\}h(\bq),
  \qquad \left(q>\frac{\pi}{L_p}\right)
\end{equation}
\end{widetext}
where the function $\Lambda_x$ depends on the wave vector, the strain
rate $\dot\gamma$, the kinetic coefficient $\Lambda(\bq)$, and the
coarse-graining length. The non-linear advection term, which arises
from assuming an affine distortion of the membrane, induces the
non-trivial renormalization of the dynamics under coarse-graining.
Hence, the ``tension'', as such, is given by 
\begin{equation}
  \label{eq:2}
  \sigma=\Lambda_x
\Lambda^{-1}(\bq)
\end{equation}
and is a function of the strain rate and is
typically wave vector dependent.

Our calculation is similar to renormalization group calculations of
the anisotropic Burgers equation, studied in the context of sandpiles
\cite{HwaK92}, and of phase separating systems under shear flow
\cite{bray01a,bray01b}. However, unlike the calculation of critical
behavior, in which coarse-graining proceeds until scaling is found,
coarse-graining in our case can only be performed up to the natural
physical cutoff corresponding to the collision length $L_p$.  This may
or may not be in the scaling regime; if it is not, then there are
generally other terms generated by the coarse-graining procedure.
However, such a calculation is generally impossible, so instead we
estimate the coarse-grained dynamics from that obtained by a
non-trivial scaling fixed point calculation. Moreover, we expect that
terms that eventually become irrelevant in the scaling regime are in
the process of being driven close to zero at the true coarse-graining
length $L_p$, and in any case are of higher wave vector and hence not
present in a hydrodynamic description.

Our one loop perturbation expansion of the dynamics yields, for any
$\Lambda(\boldsymbol{q})$, a tension that scales as $\sigma\sim
\gd^2$. By considering the energetic cost involved in bending and
stretching a single membrane, Zilman and Granek \cite{zilman99}
estimated a tension (isotropic) with the same scaling.  However, upon
performing a coarse-graining one finds a dependence $\sigma\sim
\dot{\gamma}^{\varepsilon}$, where $\varepsilon$ depends on
$\Lambda(\boldsymbol{q})$.
  
In Section~\ref{sec:equations} we summarize single membrane relaxation
dynamics, discuss previous studies of the dynamics of $h$ and 
the coarse-grained smectic displacement variable $u$ in shear flow
and derive the equation of motion for a membrane in shear in the
$c$ orientation. In Section~\ref{sec:analysis} we outline the 
coarse-graining procedure
and extract results for the renormalized dynamics, for different
relaxation mechanisms. We conclude in
Section~\ref{sec:disc-flows-coarse} with a discussion. The appendices
collect calculations of the permeation dynamics and the details of the
renormalization calculation.
\section{Membrane Dynamics}\label{sec:equations}
\subsection{Membrane Dynamics without Shear Flow}
First we review the equilibrium properties of a Helfrich-stabilized
lamellar phase, in preparation for studying its dynamics in flow.  The
long-ranged entropic interaction, characteristic of such a phase, is a
consequence of colliding membranes due to thermal fluctuations.  A key
notion is the characteristic distance $L_p$ between the collisions,
marking the transition in length scale between the membrane and bulk
smectic behaviour. This enables us to relate the static behaviour of a
single membrane at a mean layer spacing $\ell$ to the compression elasticity
of the lamellar phase.

In general, if $h(\bx,t)$ is the height of a membrane above a plane in
$d$ dimensional space,
where $\bx = (x, \bx_{\perp})$ is a $(d-1)$ dimensional vector in the
plane and $t$ is time, the
elastic free energy in the Monge gauge 
($\langle (\boldsymbol \nabla h)^2 \rangle \ll 1$) 
is given by the Helfrich Hamiltonian
\cite{helfrich78},
\begin{equation}
  \label{eq:0d}
  {\cal H} = {\tfrac12}\kappa \int d^{d-1} \! x  \left(\nabla^2 
  h(\bx)\right)^2  
\end{equation}
where the bending modulus $\kappa$ has dimensions of (energy)$\ast$(length)$^{3-d}$.
For the rest of the paper we use discrete or continuous Fourier
transforms as convenient,
\begin{widetext}
\begin{equation}
  \label{eq:3}
  h(\bx ,t)=
 \begin{cases}  
   \displaystyle {\sum_{\bq }} {\sum_\omega} h(\bq ,\omega)
   e^{i(\bq \cdot \bx - \omega t)}& \text{(discrete)} \\[24truept]
   \displaystyle \int_{-\infty}^{+\infty}
   \frac{d\omega}{2\pi}\int_{\bq } \frac{d^{d-1}\bq}{(2\pi)^{d-1}}
   h(\bq ,\omega)
   e^{i(\bq \cdot \bx - \omega t)}& \text{(continous),}\\
 \end{cases}
\end{equation}
\end{widetext}
where $\bq = q_x \hat{x} + \bqp \cdot \hat{\bx}_{\perp}$ and $\omega$
is the frequency.
For clarity we reproduce results in the rest of this section 
for $d=3$ so that $\bx_{\perp}=y\hat {\boldsymbol y}$.
In terms of the Fourier components $h(\bq)$, the equipartition theorem
gives the equilibrium height correlation function $\langle |h(\bq)|^2
\rangle =T/\kappa |\bq|^4$, where we take the Boltzmann constant $k_B=1$.
The statistics of small fluctuations with
wave vectors $qL_p>1$ are adequately determined by the Helfrich
Hamiltonian.  However, the behaviour of large fluctuations with
wave vectors $qL_p<1$, is complicated by the steric repulsion.
Constraining the fluctuations to within a layer spacing determines the
mean transverse length $L_p$ between collisions, which scales as
$L_p\sim \ell\sqrt{\kappa/T}$. The loss of entropy associated with each
collision contributes to the free energy change of the lamellar stack
under a change of layer spacing, and hence a bulk compression modulus
that scales as $B\sim T^2/\kappa \ell^3$ \cite{helfrich78}.  Long
wavelength ($qL_P<1$) lamellar behaviour is described in terms of the
displacement $u$ of the mean layer position from its mean value,
defined by
\begin{equation}
  \label{eq:61}
u(x,y,z=n\ell)=\int_{A(L_p)} 
\left[h(x'-x,y'-y) - n\ell\right]\,\frac{d^2x'}{A(L_p)},
\end{equation}
where $A(L_p)$ is the typical membrane area per collision.
Bulk equilibrium smectic behavior can be obtained by removing
the small wavelength, high $q$, height degrees of freedom $h$, until a
coarse-grained description entirely in terms of the long 
wavelength variable $u$
remains \cite{GolubovicL89}.  Our goal is to carry out this procedure
for the dynamics of a single fluctuating membrane in flow, by
coarse-graining the dynamical description up to the collision length.
A complete calculation would naturally need to simultaneously
incorporate the steric repulsion.

The relaxation dynamics of a single membrane can be described by a
Langevin equation,
\begin{equation}
  \label{eq:41}
  \partial_t h(\bx,t) = -\int d^2x'\Lambda \left(\bx-\bx'\right)
  \frac{\delta \cal{H}} {\delta h(\bx',t)} + \xi(\bx,t) ,
\end{equation}
where the thermal noise $\xi(\bq ,t)$ describes the
neglected microscopic degrees of freedom. The noise has zero mean and
a variance given by the Fluctuation Dissipation Theorem,
 \begin{equation}
  \label{eq:62b}
  \langle \xi(\bq_1,t_1)\xi(\bq_2,t_2)\rangle = 
  2T\Lambda(\bq_1 )\delta\left(\bq_1+\bq_2\right)\delta\left(t_1-t_2\right) .
\end{equation}

A spatial Fourier transformation gives
\begin{equation}
  \label{eq:42}
  \partial_t h(\bq,t) = -\Lambda(\bq)\frac{\delta \cal{H}} {\delta
    h(-\bq)} + \xi(\bq,t).
\end{equation}
The kinetic function $\Lambda\left(\boldsymbol{q}\right)$ depends on the
details of the fluid-membrane coupling; a general form is
\begin{equation}
 \begin{split}
  \label{eq:47}
  \Lambda(\bq)& =\Lambda_0 |\bq|^{m} \\
  & \sim\eta^{-1} l^{m+1} |\bq|^m,
 \end{split}
\end{equation}
where the exponent $m$ and the associated length scale $l$ depend on
the relaxation mechanism.

Three relaxation mechanisms are summarized in Fig.~\ref{fig:0a} with
relaxation functions given by
\begin{equation}
  \Lambda(\bq) \sim 
   \begin{cases}
     \eta^{-1} q^{-1} & m=-1, \quad \: \text{isolated}\\
     \eta^{-1}\zeta q^{0}& m=0, \qquad \text{permeable}\\
     \eta^{-1} \ell^3 q^2 & m=2, \qquad \text{confined fluid}.
  \end{cases}  \label{eq:45}
\end{equation} 
Thin, impermeable membranes exhibit two regimes: (i) ($m=-1$) For
$q^{-1}>\ell$ the membrane is damped by viscous solvent drag.  Solving
the linearized Navier-Stokes equations for the solvent flow yields
$\Lambda(\bq)=1/4\eta q$ \cite{Broch-Lenn75}, whence $i\omega=\kappa
q^3 /4\eta$.  (ii) ($m=2$) For $q^{-1}<\ell$, solvent flow is screened by
the surrounding membranes, leading to $i\omega = \kappa \ell^3 q^6
/16\eta$ \cite{MessagerBP90}.  Permeable membranes exhibit an
additional regime.  (iii) ($m=0$) For $\zeta<q^{-1}<\ell$ where $\zeta$
is a permeation length scale that depends on the size and the density
of microscopic defects and membrane thickness, $\Lambda(\bq)\sim
\zeta/\eta$. A simple model of cylindrical pores (common in lamellar
phases) of width $w$ and mean separation $R$ leads to $\zeta\sim R^2
\tau/w^4$, where $\tau$ is the thickness of the membrane (see
Appendix~\ref{sec:perm-length-scale}). We may envisage 
particular systems in which, at sufficiently high permeabilities
$\zeta^{-1}<\ell^{-1}$ the isolated regime is excluded, and others for
which $\zeta^{-1}<L_p^{-1}$ so that only the permeable regime remains.
\begin{widetext}
  \begin{center}
    \begin{figure}
      \includegraphics[width=14cm]{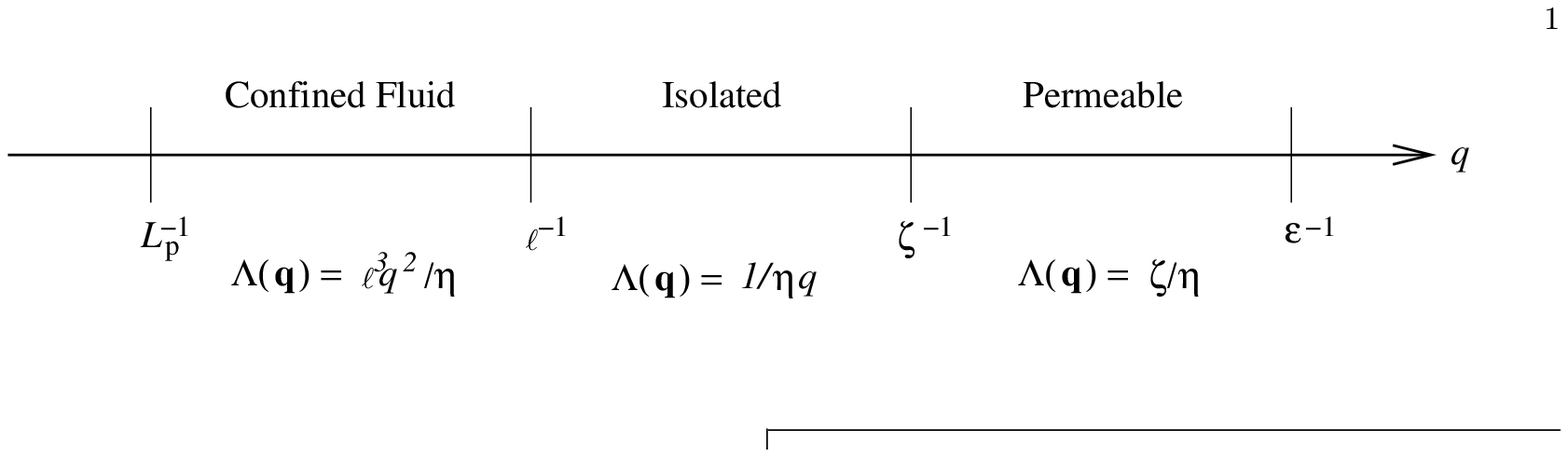}
      \caption{Different scaling regimes for the kinetic
        coefficient. $\zeta$ is the permeation length and $\varepsilon$ is
        the length at which the dynamical description breaks down.}
      \label{fig:0a}
    \end{figure}
  \end{center}  
\end{widetext}
\subsection{Equations of Motion in Shear Flow}
\subsubsection{Previous studies: linear advection}
Before we consider the dynamics of a single fluctuating membrane in
flow in the $c$ orientation, we review the equations of motion
previously used to describe lamellar phases in flow.  Milner and
Goulian \cite{goulian95} explored the stability with respect to $a$
and $c$ orientations of a thermotropic smectic under flow.  Since the
layer fluctuations of most thermotropic systems are small relative to
the length scale of the layer spacing, shear will not have a
significant effect on the shape of the layers or the internal
structure.  Thus the appropriate parametrization is entirely in terms
of the broken symmetry smectic displacement variable $u(\bf r)$.  In
flow, this quantity simply advects as a passive scalar. In simple
shear flow with average flow velocity parallel to $\hat{\bx}$, and
velocity gradient direction $\hat{\boldsymbol n}$, the flow field 
is
\begin{equation}
 \begin{split}
  \label{eq:44}
  {\boldsymbol v}({\boldsymbol r}) &= ({\boldsymbol r} \cdot
  {\hat{\boldsymbol n}}) \gd \hat{\bx} \\
  \hat{\boldsymbol n} &= \cos \theta \hat{\boldsymbol z} +\sin \theta
  {\hat{\boldsymbol y}}.\\
 \end{split}
\end{equation}
where 
$\hat{\boldsymbol n}=\hat{\boldsymbol y}$ in the $a$ orientation and 
$\hat{\boldsymbol n}=\hat{\boldsymbol z}$ in the $c$ orientation. In
this case, the dynamics of the smectic phase is
\begin{equation}
  \label{eq:43}
  \left(\partial_t - \gd q_x \hat{\boldsymbol n} \cdot 
  \frac{\partial}{\partial \bq} \right)u(q,t) = 
  -\beta(q)\frac{\delta \cal{F}}{\delta u(q,t)} + \chi(\bq ,t)  
\end{equation}
where $\cal{F}$ is the smectic free energy, $\chi(\bx,t)$ is the noise
and $\beta(q)$ is the smectic kinetic coefficient.  After minimising
an effective free energy expanded in powers of strain rate, Milner and
Goulian showed that the $a$ orientation is stable with respect to the
$c$ orientation; moreover, they demonstrated that a more rigorous
calculation of the dynamic response function incorporating non-linear
terms in the free energy exhibits the same result.

Earlier, Bruinsma and Rabin \cite{bruinsmarabin92} studied
the effect of shear on a lyotropic smectic phase in the $c$
orientation. In most of their calculations they assumed 
that flow doesn't change the membrane shape and character, and hence 
considered the same passive advection of $u$ when calculating the
effect of shear on the smectic hydrodynamic dispersion relations.
However, they did consider the degree to which flow influences the
microscopic height variable $h$, and estimated the shear rate
$\dot{\gamma}_c$ at which an affine deformation of individual
membranes leads to significant suppression of fluctuations,
\begin{equation}
  \label{eq:51}
  \gd_c =\frac{T^{5/2}}{\kappa^{3/2}\eta \ell^3}.
\end{equation}
This was obtained using, effectively,
the isolated impermeable membrane approximation
$(m=0)$ for $\Lambda(\boldsymbol{q})$.  

Finally, Ramaswamy predicted that a dilute Helfrich-stabilized
lamellar phase collapses when subjected to a flow field ${\boldsymbol
  v}=\gd y \hat{\bx}$ in the $a$ orientation \cite{Ramaswamy92b}. He
considered the motion of a membrane in shear, and argued that
only the confined fluid relaxation mode $(m=2)$ is relevant for
wavelengths less than $L_p$\footnote{Since the range of wave vectors
  from $d^{-1}$ to $L_p^{-1}\sim[d\sqrt{\kappa/T}]^{-1}$ is often very
  small or non-existent in a lamellar phase, general applicability of
  this theory is unclear.}:
\begin{equation}
  \label{eq:46}
  \partial_t h+\gd y \partial_x h = 
  \Lambda_0 \nabla^2 \frac{\delta \cal{H}}{\delta h} + \xi(\bx,t).
\end{equation}
The linearity of the advection term leads to an analytic form for the
height correlation that decreases with shear. 
Ramaswamy demonstrated that at a critical shear rate
\begin{equation}
  \label{eq:49}
  \gd_c=\frac{T^3}{\kappa^2 \eta \ell^3}
\end{equation}
the fluctuations are significantly suppressed, provoking a layer
collapse. Experimental confirmation of this was subsequently reported
by Alkahwaji and Kellay \cite{alkahwaji00}. 
Many other studies have incorporated the simple advection term in the
dynamics for $u$: for example, Cates and Milner studied the effect of
flow on the isotropic-lamellar transition \cite{catesmilner89}, and
Fredrickson studied the effect of flow on diblock lamellar phases
\cite{Fredrickson86}.  All of these treatments are suitable if flow
does not significantly perturb the layer microstructure. 
In contrast the perturbation of the microstructure in
thermotropic lamellar phases was studied in
Ref.~\cite{AuernhammerBP00}. In this case, shear flow was shown to
introduce a tilt in the layers of a smectic-A liquid crystal.
This led to layer reorientation and the possibility of an instability.
\subsubsection{Membrane in Flow in the $c$ orientation; non-linear advection}
Now we consider the effect of a shear field ${\boldsymbol v}=\gd z
\hat{\bx}$ on a fluctuating membrane in the lamellar phase in the $c$
orientation (Fig.~\ref{fig:0}).  If we affinely transform the membrane
in a small time $\delta t$, the height field advects according to
\begin{widetext}
  \begin{center}
    \begin{figure}[htbp]
      \includegraphics[width=14cm]{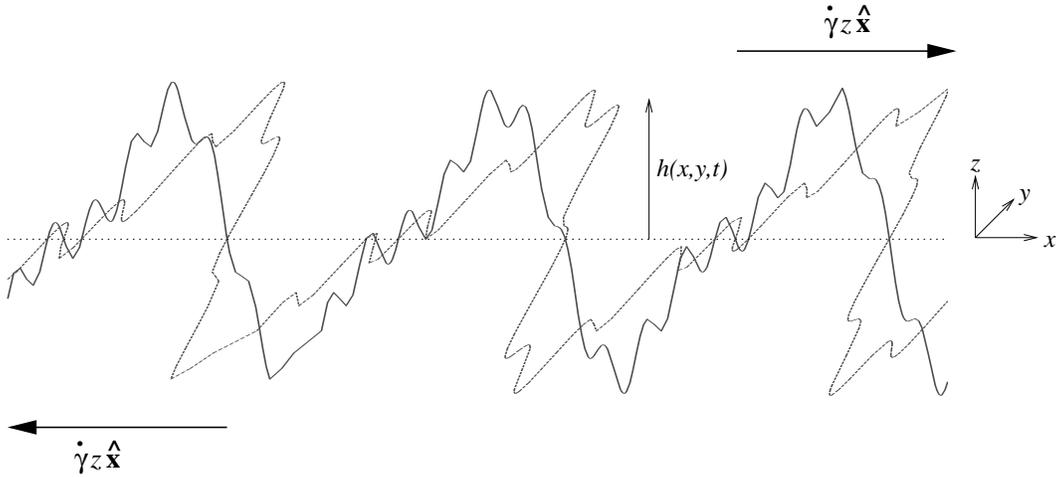}
      \caption{A fluctuating membrane at rest (solid line) and 
      subject to a shear field (dashed line).}
      \label{fig:0}
    \end{figure}
  \end{center}  
\begin{align}
\label{eq:50}
  h(x,y, t+\delta t) &= h(x-\gd z \delta t, y, t)&
  \Longrightarrow &&\frac{\partial h}{\partial t} \delta t & =
  \frac{\partial h}{\partial x}(-\gd z \delta t) + {\cal O}(\delta
  t)^2 .
\end{align}
The key point is that, because shear is assumed to advect the
membrane, $z$ should be equal to the height field $h$. 
This leads to a non-linear advective term, so that the
equation of motion for a membrane in $(d-1)$ dimensions becomes:
\begin{subequations}
  \label{eq:1}
 \begin{align}
  \left[\partial_t+
    \dot{\gamma}h\left(\bx ,t\right)\frac{\partial} {\partial
      x}\right] h(\bx,t) = 
   -\int d^{d-1}x'\Lambda \left(\bx-\bx'\right)
  \frac{\delta \cal{H}} {\delta h(\bx',t)} + \xi(\bx,t) \\
  -i\omega h(\bq, \om ) +i \gd \sum_{\Omega} 
  \sum_{\bk}(q_x -k_x ) h(\bk , \Om) h(\bq - \bk , \om - \Om) 
  = -\Lambda(\bq)\kappa q^4\, h(\bq) +\xi (\bq, \om). \label{eq:14}
 \end{align}
\end{subequations}
\end{widetext}
The advective term stretches the membrane, leading to a restoring
``tension''.  In contrast, the relevant advective term for a
thermotropic in the $c$ orientation remains $y\partial u/\partial x$,
because the perturbation of the membrane structure is negligible, and
undulations of $u$ are always presumed to be of much smaller wavelength
than the layer spacing.  The noise has variance
\begin{subequations}
  \label{eq:8}
\begin{align}
  \langle \xi(\bx_1,t_1)\xi(\bx_2,t_2)\rangle & = 
  D(\boldsymbol{x}_1-\boldsymbol{x}_2) \delta \left(\bx_1-\bx_2\right)
  \delta \left(t_1-t_2\right)\\
\langle \xi (\bq_1 , \om_1 ) \xi (\bq_2,\om_2 ) \rangle 
& = D(\bq_1)\delta^{d-1}(\bq_1 + \bq_2 ) \delta(\om_1 + \om_2 )\label{eq:8b}\\[10truept]
D(\bq)& =2 T\Lambda(\boldsymbol{q}) = 2T\Lambda_0 |\bq|^{m}\label{eq:52}\\
& \equiv D_0 |\bq|^{m}, \label{eq:37} 
\end{align}
\end{subequations}
where Eq.~(\ref{eq:52}) is the Fluctuation Dissipation Theorem.

A similar equation was derived for a diffuse interface between phase
separated domains subject to shear by Bray and co-workers 
\cite{bray01a,bray01b}.  However,
there are two important differences.  First, the ``interface'' of a
membrane in the lamellar phase is intrinsically sharp. Second,
although both bending ($\sim q^4$) and surface ($\sim q^2$) energies
are present for a diffuse interface, at long length scales only the
latter is relevant.  On the other hand, surface tension is not present
\textit{a priori} in equilibrium lamellar phases, so the Helfrich
Hamiltonian is dominated by the bending energy.

For no flow $\gd =0$, the non-linear term in Eq.~(\ref{eq:14})
vanishes and
\begin{equation}
  \label{eq:29}
  h (\bq , \om) = G(\bq , \om) \xi(\bq , \om).
\end{equation}
This defines the bare linear propagator, equivalent to the linearized
equation of motion,
\begin{equation}
\label{eq:29b}
  G^{-1}(\bq ,\om)=-i\om + \Lambda(\bq)  S^{-1}(\bq),
\end{equation}
where $S^{-1}(\bq) = \kappa |\bq|^4$.
The poles of $G(\boldsymbol{q},\omega)$ yield the dispersion relation
for the decay times of height fluctuations.  The equal time height
correlation function may be calculated from
\begin{subequations}
\begin{align}
  C(\bq)&=\int_{-\infty}^{\infty}\frac{d\omega}{2\pi}   C(\bq , \om)\\
  &=\int_{-\infty}^{\infty}\frac{d\omega}{2\pi} \langle h^* (\bq ,\om )h(\bq ,\om ) \rangle   \label{eq:26}\\
  &={T}S(\bm{q}).
\end{align}
\end{subequations}
Equations (\ref{eq:14}, \ref{eq:8b}, \ref{eq:37}) define our model, with
$\Lambda(\boldsymbol{q})$ given by Eq.~(\ref{eq:45}). In the next
section we examine the effect of the non-linear advective coupling on
the spectrum of height fluctuations, through its effect on the linear
propagator. 
\section{Coarse-Graining Procedure}\label{sec:analysis}
\subsection{Description of the problem}
We wish to calculate the effective dynamical response of the height
field, determined by Eq.~(\ref{eq:14}), for different models specified
by the relaxation function $\Lambda(\boldsymbol{q})$ and hence $m$.
In Section \ref{sec:one-loop-correction} below we show that a one-loop
perturbation analysis leads to a divergent response for all candidate
values for $m$. In order to avoid this unphysical divergence we apply
a Renormalization Group (RG) procedure in Section
\ref{sec:renorm-group-rg} to coarse-grain the system by removing the
small scale and faster degrees of freedom, leaving an effective long
wavelength theory. This procedure naturally generates a dynamic
response that scales as $q_x^2$ due to the advective non-linearity,
is suggestive of a tension, and restores non-singular behaviour.

In the study of dynamical critical phenomena \cite{hohenberg77} the
description of a system is coarse-grained until fixed points are
found for which the system is self similar. Hence, a condition is
found on the parameters of the dynamical equations of motion that
leaves the dynamics of the height field, scale invariant.  Contrary to
a system at its critical point, the lamellar system here can only be
coarse-grained up to the natural physical cutoff of the collision
length.  However, since this length is not necessarily in the scaling
regime, such a calculation requires detailed knowledge of the 
flow equations for all new terms generated in the equation of motion,
which is generally impossible.  On the other hand, if the collision
length is close to, or larger than, the wavelength at which the
scaling regime applies, we may use the simpler fixed point calculation
as a guideline to estimate the effective long wavelength dynamics. In
fact, even if we are not in the scaling regime, the other terms
generated by coarse-graining are of higher order in wave vector and are
irrelevant at the fixed point, and are thus in the process of being
driven to zero during the coarse-graining procedure. In the
hydrodynamic limit such terms are neglected.

In the following analysis, we
coarse-grain the system to generate the lowest order (in wave vector)
additional term in the equation of motion, which depends on the
coarse-graining length $L_p$ and the particular 
relaxation mechanism $\Lambda(\boldsymbol{q})$, parametrized by $m$.
However, unless the appropriate
range for the length scales of a given relaxation mechanism (see
Fig.~\ref{fig:0a}) encompasses the entire coarse-graining range, a
full calculation demands a more precise choice of
$\Lambda(\boldsymbol{q})$. We shall not attempt
such a calculation, but give the results under the assumption that a 
single value of $m$ applies throughout the wave vector regime
of interest.

A consequence of the particular energy ($\sim q^2$) intrinsic to a
diffuse interface is that Bray and co-workers were able to extract
results only for ``models'' $m\geq 0$. In comparison, for a
fluctuating membrane ($\sim q^4$), we may examine the relaxation
mechanisms $m\geq-2$ including the case of an isolated impermeable
membrane that relaxes by the hydrodynamic interaction of the
surrounding solvent.
\subsection{One-loop correction to  $G(\bq , \om)$}\label{sec:one-loop-correction}
We start by rewriting Eq.~(\ref{eq:14}) as
\begin{widetext}
\begin{equation}
  \label{eq:6}
  h(\bq ,\om )= G(\bq,\om)\left[\xi (\bq, \om) - i \gd \sum_{\Omega} 
  \sum_{\bk}(q_x -k_x ) h(\bk , \Om) h(\bq - \bk , \om - \Om) \right] .
\end{equation}
Eq.~(\ref{eq:6}) is shown in Fig.~\ref{fig:3}(a).
\begin{figure}[tbp]
     \begin{center}
       \includegraphics[width=12cm]{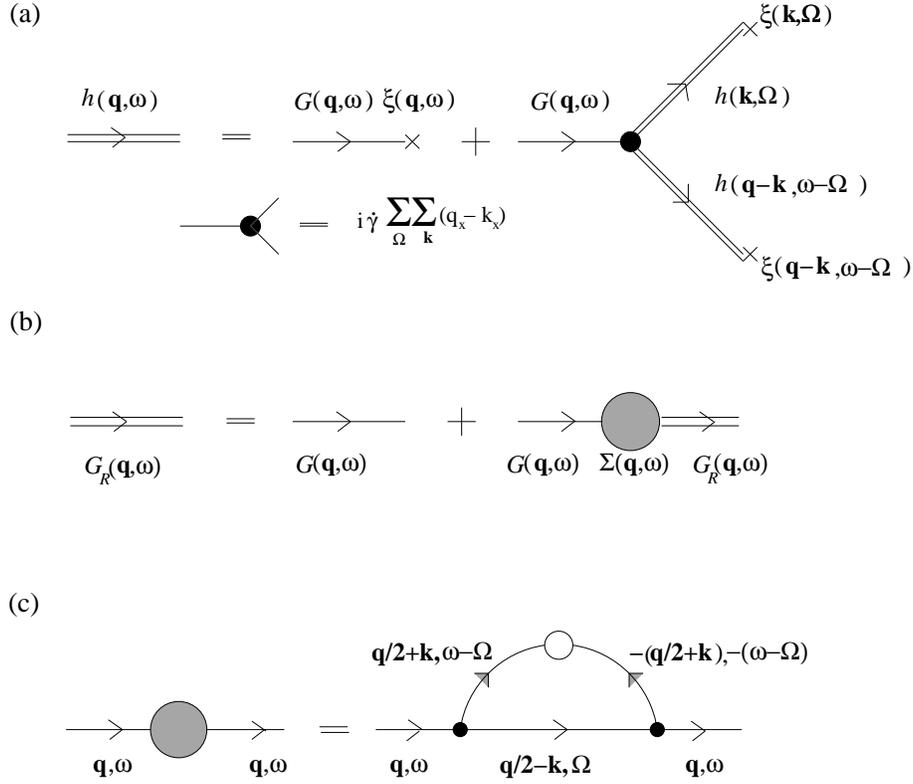}
     \caption{(a) Diagrammatic representation of
       Eq.~(\protect{\ref{eq:6}}). Solid circles $\bullet$ represent
       non-linear vertices.  (b) Dyson Equation for the renormalized
       propagator in terms of the bare propagator (single lines) and
       the self-energy (shaded circle). (c) One-loop correction to the
       self-energy. Open circles $\circ$ represent noise
       contractions, $\langle\xi\xi\rangle$.}
      \label{fig:3}
     \end{center}  
\end{figure}
Adding a perturbation $f(\bq,\om)$ to Eq.~(\ref{eq:6}) and averaging
over the stochastic noise $\xi$ results in a renormalized linear
propagator $G_R$, defined by
\begin{equation}
  \label{eq:60}
G_R(\bq, \om ) =\lim_{f\to 0}  \frac{\partial\langle h(\bq , \om )\rangle}{\partial
f(\bq , \om )},
\end{equation}
given by the Dyson equation (Fig.~\ref{fig:3}b) \cite{ma1976},
\begin{align}
  G_R(\bq , \om )^{-1}& = G(\bq, \om )^{-1} -
  \Sigma(\bq , \om) .\label{eq:12} 
\end{align}
The self energy $\Sigma(\bq, \om)$ shifts the poles of the 
effective propagator $G_R$ and equivalently renormalizes the
effective equation of motion (Eq.~\ref{eq:29b}) 
for the long wavelength degrees of freedom.  
The one-loop contribution to the self energy
$\Sigma( \bq, \om )$ (Fig.~\ref{fig:3}c) yields
\begin{equation}
 \begin{split}
  \label{eq:13c}
  \Sigma(\bq , \om )= -\gd^2 \sum_{\Om } \sum_{\bk }
  (q_x- k_x) G(\bk , \Om) & G(\bq - \bk , \om - \Om) \\
  & \times \Big[(q_x -k_x)G(-\bk , -\Om )D(\bk) + k_x G(\bk - \bq ,
  \Om - \om )D(\bk-\bq)\Big].
 \end{split} 
\end{equation}
A change of co-ordinates 
$\bk\to\bq /2+ \bk $ and $\bk \to \bq /2-\bk$ in the first and second
terms respectively of Eq.~(\ref{eq:13c}) gives
\begin{equation}
  \label{eq:13d}
  \Sigma(\bq , \om )= -\gd^2 \sum_{\Om } \sum_{\bk }
  q_x \left(\frac{q_x}{2}-k_x \right) G\left(\frac{\bq}{2}-\bk ,\Om \right) 
   \left|G \left(\frac{\bq}{2}+\bk ,\om-\Om \right) \right|^2 
   D_0\left|\frac{\bq}{2}+\bk\right|^m,
\end{equation}
shown in Fig.~\ref{fig:3}.  On symmetry grounds, the self energy can
be written as
\begin{equation}
 \label{eq:13f}
  \Sigma(\bq , \om )= a_2(\bm{q},\omega) q_x^2  + a_4(\bm{q},\omega)
  q_x^4  + \ldots,
\end{equation}
in which case the renormalized propagator is given by 
\begin{equation}
  \label{eq:63}
G_R(\bq , \om )^{-1}\simeq
-i\om + \Lambda(\bq)  S^{-1}(\bq) + 
a_2(\bm{q},\omega) q_x^2.
\end{equation}
and the associated noise correlation is 
\begin{equation}
  \label{eq:62}
  \langle \xi (\bq_1 , \om_1 ) \xi (\bq_2 ,\om_2 ) \rangle 
= \big(D(\bq_1) + D_x q_x^2 \big)\delta(\bq_1 + \bq_2 ) 
\delta(\om_1 + \om_2 ). 
\end{equation}
The effect of fluctuations on the slow hydrodynamic regime is due to
the leading long wavelength static behavior, $\omega=0$.  Hence, we
expand to order $q_x^2$ for $\omega\rightarrow 0$. Note that we
implicitly ignore renormalizations of the frequency dependence, which
should be sufficient to obtain scaling relations.

After converting the sum to a continuum integral
and integrating out the frequency $\Omega$ (Eq.~\ref{eq:75c} 
in Appendix~\ref{sec:self-energy}), the 
lowest order term in $q_x$ that appears in the static self-energy is
\begin{equation}
  \label{eq:13e}
 \Sigma(\bq ,0)= -\gd^2 q_x^2 \frac{D_0}{2(\kappa\Lambda_0)^2}  
  \int_{0}^{\infty}\frac{k^{d-2}dk S_{d-1}}{(2\pi)^{d-1}} 
  \left[\frac{1}{4|\bk|^{8+m}}+\frac{2k_x(k_x^3+k_x
      k_{\perp}^2)}{2|\bk|^{12+m}}\right] + {\cal O\/}(q_x^4),
\end{equation}
\end{widetext}
where $S_{d-1}$ is the unit sphere surface area in $(d-1)$ dimensions.
For $10+m-d>0$, which encompasses all of our relaxation regimes
$m=-1,0,2$, this integral diverges.  In fact, before changing the
limits of integration in Eq.~(\ref{eq:13c}), the divergence in
Eq.~(\ref{eq:13c}) can be traced to the second term in square brackets
at $\bq =\bk $.  That is, the divergence is actually independent of
the lower cutoff. Such a singular response suggests that a description
in terms of the microscopic degrees of freedom is physically
inconsistent.  To proceed, we note that a tension-like term $\Lambda_x
q_x^2$ added to the propagator yields non-singular behavior
(Eq.~\ref{eq:48} in Appendix~\ref{sec:self-energy}).  Furthermore, we
will show below that the gradual thinning of degrees of freedom during
coarse-graining directly generates terms that render the response
physical (Eq.~\ref{eq:31b} in Appendix~\ref{sec:self-energy}). Hence,
we argue that the correct physical description of the highly
fluctuating system in shear must formally be derived by projecting out
the small scale degrees of freedom. We will outline this procedure
next.
\subsection{Renormalization Group (RG) analysis}\label{sec:renorm-group-rg}
Our goal is to successively ``integrate out'' the small scale fast
degrees of freedom from the equation of motion, to yield an effective
equation of motion for the remaining long wavelength degrees of
freedom \cite{onuki79}. Schematically, we can write 
\begin{equation}
  \label{eq:64}
  h(\bx ,t) =
  h^> (\bx ,t)  +   h^<(\bx ,t), 
\end{equation}
where $h^<(\bx ,t)$ and $h^>(\bx ,t)$ are, respectively, small and
large wave vector degrees of freedom. Upon removing the faster $h^>(\bx
,t)$, an effective equation of motion for $h^<(\bx ,t)$ will be
generated. The form of the equation will differ from the original
equation because of the non-linear advective term, which couples
different modes together. Note that, generally, there are also
non-linear terms (of order $h^3$) due to deviations from the Monge
gauge limit and higher order advection terms, but the restriction of
transverse length scales to within a patch length ensures that we
remain close to this limit. The new equation of motion is most easily
cast in terms of the renormalized propagator $G_R^<$ of the long
wavelength degrees of freedom.  As noted above, we expect the
contributions to the propagator, and generally to the noise, to be
even powers in $q_x^2$, because of the symmetry of the non-linearity.
In the hydrodynamic and long wavelength limit, and indeed because
higher order terms are irrelevant at the non-trivial fixed point, we
focus on corrections of order $q_x^2$.

Essentially there are three steps to a momentum-shell RG in which, for
convenience, we impose a short wavelength ultraviolet cutoff $\lambda$
in the $x$-direction only; these are the fluctuations directly
suppressed by flow. Physically, the original cutoff is
$\lambda=\pi/a$, where $a$ is of order a molecular (surfactant) size,
and we are interested in coarse-graining from this length to some,
unknown, length, at which non-trivial scaling behavior is seen. The
steps are as follows:
\begin{enumerate}
\item The first step is to divide the Brillouin zone $k \in
  [0,\lambda]$ into two parts: high wave vectors $k^> \in[\lambda/b,
  \lambda]$ to be removed, and the remaining long wavelengths $k^< \in
  [0, \lambda/b]$. The elimination of (assumed) fast modes results in
  an effective renormalized propagator $G_R^<(\bq , \om)$.  Since
  there are no singularities in this range of integration, only finite
  corrections to the parameters result.
\item After coarse-graining, the resulting equation has a cutoff
  $\lambda/b$. This difference from the original model is removed by
  rescaling the length scales, $x$, $x_{\perp}$, $h$ and the time
  scale $t$.
\item Finally we look for the fixed points of the recursion relation
  at which the theory is invariant under the first two steps.
\end{enumerate}

This procedure generates a recursion equation that may be used to find
the behavior of the system in a scaling regime. Generally, this
scaling regime corresponds to a description of the system at
wavelengths longer than that wavelength at which the dynamics has
effectively been driven to the fixed point. We will use this fixed
point as an \textit{estimate} of the dynamics of the system at
wavelengths larger than the collision length $L_p$.

We follow Bray \textsl{et al.'s} study \cite{bray01a,bray01b} of the
influence of shear flow on interfacial dynamics in a phase separating
system, which is governed by square gradient terms rather than quartic
energy possessed by membranes. The scale transformation takes the form
\begin{equation}
  \label{eq:15}
  x=bx',\qquad \bx_{\perp} =b^\zeta \bx'_{\perp}, \qquad
  h=b^{\chi}h', \qquad t=b^z t'.
\end{equation}
Since shear suppresses the fluctuations in the $x$-direction we expect
to find $\zeta\leq 1$ when the shear is relevant.  We will see that
condition is only satisfied if $m\geq-2$.  Since we consider only
models $m\geq -1$, this condition is always satisfied.  In such cases
the transverse part $\bqp$ dominates $q_x$ in the terms involving
powers of $|\bq|$ so that the bare propagator is renormalized to
\begin{equation}
  \label{eq:16}
  G_R^{-1}(\bq, \om)=-i\om + \Lambda(\bqp) S^{-1} (\bqp) + \Lambda_x q_x^2,
\end{equation}
and the noise correlator is
\begin{widetext}
\begin{equation}
  \langle \xi (\bq_1 , \om_1 ) \xi (\bq_2 ,\om_2 ) \rangle 
= \big(D({{\boldsymbol q}_1}_{\!\perp}) + D_x q_x^2 \big)\delta(\bq_1 + \bq_2 ) \delta(\om_1 + \om_2 ). 
  \label{eq:17}
\end{equation}
\end{widetext}
We have included the lowest order correction to $G_R$ and the noise
from the non-linearity.
Applying the rescaling Eq.~(\ref{eq:15}) yields
rescaled parameters in the equation of motion and the noise
correlator:
\begin{subequations}
\label{eq:scaling}
\begin{align}
  \gd' &=b^{\chi+z-1}\gd  \label{eq:18} \\
  \Lambda' &=b^{z-(4+m)\zeta} \Lambda  \label{eq:19} \\
  \Lambda'_x &=b^{z-2} \Lambda_x + \ldots  \label{eq:20} \\
  D'_0 &=b^{z-2\chi -1-m\zeta-(d-2)}D_0  \label{eq:21} \\
  D'_x &=b^{z-2\chi -3-(d-2)\zeta}D_x + \ldots \label{eq:22}
\end{align}
\end{subequations}
The parameters $\Lambda'_x$ and $D'_x$ acquire perturbative
corrections due to the coarse-graining step of the RG procedure. In
contrast $\gd$, $\Lambda$ and $D_0$ do not acquire perturbative
corrections.  The nonrenormalizability of $\gd$ follows from Galilean
invariance of Eq.~(\ref{eq:50}), which 
transforms $t\to t + \delta t$ and $x \to x-\gd h \delta t$ 
in the equation of motion.

We first examine the linear theory to identify the critical dimension
$d_c$. Since there are no perturbative corrections to a linear theory,
the requirement that the exponents for $b$ vanish in
Eqs.~(\ref{eq:19}-\ref{eq:22}) yields
the conditions
\begin{equation}
  \label{eq:23}
  z_0=2, \qquad \zeta_0=\frac{2}{m+4}, \qquad 
  \chi_0=\frac{8-m-2d}{2(m+4)}.
\end{equation}
The subscripts denote the application to the linear theory.
Eq.~(\ref{eq:18}) determines the relevance of the shear rate $\gd$ on
the coarse-graining, at the trivial fixed point. From
Eq.~(\ref{eq:23}) we obtain $\chi_0 + z_0 -1 =
(m-2d+16)/\left[2(4+m)\right] $. Therefore, $\gd$ is relevant for
$d<d_c$ where
\begin{equation}
  \label{eq:24}
  d_c=\frac{16+m}{2} ,\qquad m \geq -2 .
\end{equation}

\begin{figure}[htbp]
     \begin{center}
       \includegraphics[width=8.0truecm]{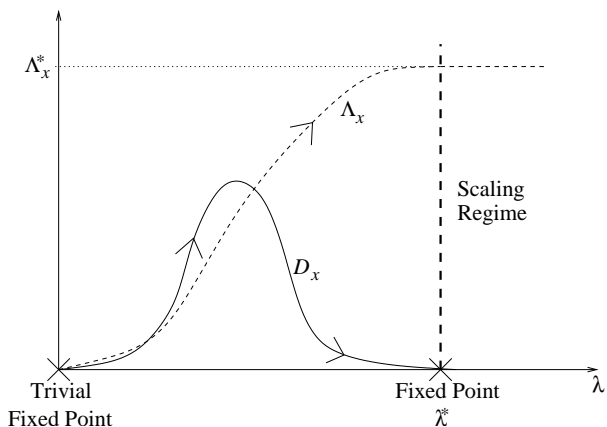}
     \caption{Schematic flows  of $\Lambda_x$ and
       $D_x$ as a function of the coarse-graining length $\lambda$.}
      \label{fig:1b}
     \end{center}  
\end{figure}

We can coarse-grain the theory perturbatively in Fourier space near
the critical dimension $d_c$ of the theory.  For $d<d_c$ we expect a
new fixed point to appear at which $\gd$, $\Lambda$ and $D_0$ are
non-zero.  Eqs.~(\ref{eq:18}, \ref{eq:19}, \ref{eq:21}) give the
corresponding exponents exactly,
\begin{widetext}
\begin{equation}
  \label{eq:25}
  z=\frac{3(4+m)}{14+2m-d}, \quad \zeta=\frac{3}{14+2m -d}, 
  \quad \chi=\frac{8-m-2d}{2(4+m)}.
\end{equation}
\end{widetext}
From Eqs.~(\ref{eq:23}) and (\ref{eq:25}) we see that $\zeta\leq1$ for
$m\geq -2$, in which case the approximation $|\bq|\sim |\bqp|$ is
consistent. From Eq.~(\ref{eq:25}) we find $D'_x=b^{-8/(4+m)}D_x$,
indicating that $D_x$ flows to zero at the fixed point 
(Fig.~\ref{fig:1b}). 

To find the values of $\Lambda_x$ and $\lambda$ at the non-trivial
fixed point we now return to the RG procedure. Integrating
Eq.~(\ref{eq:12}) over the short wavelength modes gives the equation
for the effective renormalized propagator
\begin{equation}
  \label{eq:25b}
  G_R^<(\bq , \om )^{-1}= G(\bq, \om )^{-1} - \Sigma(\bq , \om) ,
\end{equation}
where $\Sigma(\bq , \om )$ is given in Eq.~(\ref{eq:13d}).
Then, setting $b=e^l$ with $l$ infinitesimal, Eqs.~(\ref{eq:20}) and
(\ref{eq:25b}) generate a differential flow equation for $\Lambda_x$
\begin{equation}
  \label{eq:27}
  \frac{d \Lambda_x}{dl} = \Lambda_x\left[(z-2)
  -\lim_{\bq \to 0}\frac{1}{\Lambda_x q^2_x l}\Sigma(\bq ,0)\right].
\end{equation}
Step (3) of the RG analysis consists of finding the fixed points
$\Lambda^{\ast}_x$ for which the theory is invariant. This procedure
is carried out in Appendix~\ref{sec:self-energy} for each value of
$m$. Finally, we can make contact with our original discussion of an
induced ``tension'' $\sigma$, and extract a tension according to
Eq.~(\ref{eq:2}). The results are collected in Table~\ref{tab:1}.  For
$m=0$ and $m=2$ the procedure is straightforward and the results (and
$\Lambda_x$) are independent of the low-$k$ cutoff of the theory.
However, for $m=-1$ we must cut off the theory at $k=\pi/L_p$, and
hence we find a result for $\Lambda_x$ that depends on $L_p$. 

As noted in the introduction, on the basis of estimating the height
correlation function from the first term in the perturbation expansion
in Eq.~(\ref{eq:6}), we would naively expect the ``tension'' to scale
quadratically with the strain rate; this scaling was also captured by
considering the energetic cost of bending and stretching a single
membrane \cite{zilman99}. However, an anomalous scaling
$\Lambda_x\sim\dot{\gamma}^{\varepsilon_m}, \varepsilon_m\neq2$ is
generated in the scaling regime for $m=0$ and $m=2$. The case of
$m=-1$ is different. Owing to the divergence in the perpendicular
direction at the low cutoff due to the more violent fluctuations at
long wavelengths, the scaling does not follow the expected pattern
($\varepsilon_{-1}=2/3$) of the other mechanisms.  However, whether
accidentally or not, the first term in its power series satisfies
$\varepsilon_{-1}=2$ so at low strain rates it cannot be considered to
exhibit anomalous scaling; in general though, contributions from
higher order terms indicate that the scaling is also anomalous.

In addition, equating bending and tension energies leads to an
expression for the shear rate $\gd_c$ at which fluctuations are
significantly suppressed,
\begin{equation}
  \label{eq:58}
  \gd_c \sim \frac{T^{(m+3)/2}\Lambda_0}{\kappa^{(m+1)/2}{\ell}^{m+4}},
\end{equation}
where in the case $m=-1$ there are higher order contributions, shown in 
Table~\ref{tab:1}.
If we compare the critical shear rate for isolated impermeable membrane
relaxation with Bruinsma's result (Eq.~\ref{eq:51}), the scaling is the
same apart from a factor of $\sqrt{T/\kappa}$.

\begin{table*}
\begin{ruledtabular}
\caption{\label{tab:1} Row (2) shows the fixed
 points corresponding to different membrane relaxation mechanisms, given in
 row (1). Row (3) gives the ``tension'', defined by Eq.~(\ref{eq:2}) 
 as a function of the cutoff, $\lambda$
 and the relaxation mechanism. Rows (4) and (5) display the results for  
 $\lambda^{-1}\sim L_p$ and $q^{-1}\sim L_p \sim \ell\sqrt{\kappa/T}$
 respectively. Row (6) displays the critical shear rate for
 suppression of fluctuations.}
\begin{tabular}{l@{\qquad}|cccc}
& Membrane & Isolated  & Permeable  & Confined  \\
& &  ($m=-1$) &  ($m=0$) &  ($m=2$) \\
\hline \\ 
(1) & $\Lambda(\bq)=\Lambda_0 q^m$  & $\displaystyle \frac{1}{\eta q}$
& $\displaystyle \frac{\zeta q^0}{\eta}$ 
& $\displaystyle \frac{d^3q^2}{\eta}$ \\[20.0truept]
(2) & $\Lambda^*_x(\lambda, \Lambda_0)$
& $\displaystyle \alpha_0 \frac{T \lambda L_p^6}{\Lambda_0 \kappa^2}
  \gd^2 + {\cal O}(\gd^4)$
& $\displaystyle \alpha_1 \frac{T^{4/11}\Lambda_0^{3/11}}
  {\kappa^{1/11}\lambda^{10/11}}\gd^{8/11}$ 
& $\displaystyle \alpha_3 \frac{T^{2/5}\Lambda_0^{4/5}} 
  {\kappa^{1/5}\lambda^{4/5}} \gd^{4/5}$ \\ [25.0truept] 

(3) & $\sigma(\lambda , \Lambda_0, q)$ 
& $\displaystyle \alpha_0 \frac{T \lambda L_p^6}{\Lambda_0^2 \kappa^2}
  \gd^2 q + {\cal O}(\gd^4)$
& $\displaystyle \alpha_1 \frac{T^{4/11}}
  {\kappa^{1/11}\Lambda_0^{8/11}\lambda^{10/11}}\gd^{8/11}q^0$ 
& $\displaystyle \alpha_3 \frac{T^{2/5}} 
  {\kappa^{1/5}\Lambda_0^{4/5}\lambda^{4/5}} \gd^{4/5}q^{-2}$ \\ [25.0truept]
(4) & $\sigma(L_p ,q)$ 
& $\displaystyle \alpha_0 \frac{T L_p^5}{\kappa^2}\eta^2\gd^2 q+{\cal O}(\gd^4)$
& $\displaystyle \alpha_1 \frac{T^{4/11}}{\kappa^{1/11}}
   \frac{L_p^{10/11}}{\zeta^{8/11}}\eta^{8/11}\gd^{8/11}q^0$
& $\displaystyle \alpha_3 \frac{T^{2/5}}{\kappa^{1/5}}
  \frac{L_p^{2/5}}{d^{12/5}}\eta^{4/5}\gd^{4/5}q^{-2}$ \\ [25.0truept]
(5) & $\sigma(d)$
& $\displaystyle \alpha_0 \frac{\ell^4}{T}\eta^2 \gd^2+ {\cal O}(\gd^4)$
& $\displaystyle \alpha_1 \frac{\kappa^{4/11}}{T^{1/11}}
   \frac{\ell^{10/11}}{\zeta^{8/11}} \eta^{8/11}\gd^{8/11}$
& $\displaystyle \alpha_3 
\frac{\kappa^{6/5}}{T} \ell^{2/5}\eta^{4/5} \gd^{4/5}$ \\ [25.0truept]
(6) & $\gd_c$ 
& $\displaystyle \frac{T}{\eta \ell^3}\left(1-\frac12\left(\frac{T}{\eta\ell^3}\right)^2+\ldots\right)$ 
& $\displaystyle \frac{T^{3/2}\zeta}{\kappa^{1/2}\eta \ell^4}$ 
& $\displaystyle \frac{T^{5/2}}{\kappa^{3/2}\eta \ell^3}$ 
\end{tabular}
\end{ruledtabular}
\end{table*}

\subsection{One-Step Coarse-Graining}\label{sec:first-step}
So far we have demonstrated how to calculate the non-trivial scaling
behavior of the membrane, assuming that fluctuations generate a
tension-like term that renders contributions to the self-energy
non-singular. Here, we show explicitly how the ``first step'' of a
coarse-graining procedure produces such a term.  Here we coarse-grain
the system in ``one step'', by making a small perturbation to the
original microscopic cutoff $\lambda=\pi/a$, where $a$ is a typical
molecular dimension.  For a small perturbation about the trivial fixed
point ($\Lambda_x=0$) the differential flow equation Eq.~(\ref{eq:27})
becomes
\begin{equation}
  \label{eq:28}
  d\Lambda_x \simeq -\lim_{\bq \to 0}\frac{1}{q^2_x}\Sigma(\bq ,0). 
\end{equation}
We show at the end of Appendix \ref{sec:self-energy} the calculation of
the coarse-grained self-energy in the limit $\Lambda_x\to0$ 
(Eq.~\ref{eq:38}). Inserting the resulting self-energy 
(Eq.~\ref{eq:55}) into the recursion relation yields an expression
for $\Lambda_x$ and, via Eq.~(\ref{eq:2}), a ``tension'' that depends
on the cutoff and the relaxation mechanism,
\begin{equation}
  \label{eq:56}
  \sigma = \beta \frac{T}{\kappa^2\Lambda_0^2
    \lambda^{9+m-d}}\gd^2 l q^{-m} ,
\end{equation}
where $\beta$ is a numerical prefactor and $l$ is a small number that
depends on the chosen coarse-graining step. 
As expected, for all
relaxation mechanisms, a first step coarse-graining leads to a
``tension'' that is the same as the naive scaling $\sim \eta^2\gd^2$.
Thus we may infer that the coarse-graining process modifies the
dependence of the scaling for $m=0,m=2$ and also for $m=-1$ (but not
for very small strain rates).

\section{Results and Discussion}\label{sec:disc-flows-coarse}
\subsection{Single Membrane Dynamics in the Scaling Regime}
We have considered the dynamics of a single membrane in the
$c$-orientation (Fig.~\ref{fig:0b}) with respect to shear flow.
Advection couples different Fourier modes, and hence renormalizes the
effective response. We have estimated the effective long-wavelength
theory, that would be obtained by removing smaller and faster degrees
of freedom with wave vectors $q>\lambda$, by calculating the behavior
of the fluctuating membrane in the scaling regime. This dynamic
coarse-graining generates, to lowest order, a term $- \Lambda_x q_x^2$
in the long wavelength propagator. This is the principle qualitative
result of this work.

The function $\Lambda_x$
depends on wave vector $q$, the quiescent relaxation mechanism
$\Lambda(\boldsymbol q)$, the strain rate, and the wavector scale $\lambda$ to
which coarse-graining has been performed (Appendix~\ref{sec:self-energy}):
\begin{equation}
  \label{eq:7}
  \Lambda_x \sim A_m\lambda^{\delta_m}\dot{\gamma}^{\varepsilon_m}
\end{equation}
where $m$ parametrizes the relaxation mechanism
$\Lambda(\boldsymbol q)=\Lambda_0q^m$, and the constant $A_m$ depends on
$T,\kappa,L_p,$ and $\Lambda_0$.  The results are summarized in
Table~\ref{tab:1}.  This restoring term is suggestive of an
anisotropic ``tension'' $\sigma$, which would appear in the dynamics as $\sigma
q_x^2h^<(\boldsymbol q)$ (Eq.~\ref{eq:2}), 
with $\sigma = \Lambda_x/\Lambda(\boldsymbol q)$, except that the
non-analytic form generally leads to a wave vector dependence. In the
permeable limit the wavector dependence is absent, $\nu=0$, while in
other cases there is a wave vector dependence. Hence, referring to the
newly generated term as a tension is suggestive at best. Nonetheless,
this term can be expected to suppress fluctuations, and hence
influence the effective collision rate, and in turn the Helfrich
interaction potential, in the presence of shear flow. Elsewhere, we
have used an effective energetic tension to parametrize the reduction
in fluctuations and the corresponding flow-induced strain or change in
layer spacing \cite{marlowunpub}. 

It is important to recognize that, although the ``tension'' in
Table~\ref{tab:1} applies, strictly, only to wavelengths of order the
collision length, it is generated at all wavelengths larger than the
smallest cutoff and grows during the coarse-graining procedure. In
reference \cite{marlowunpub} we replaced this wavevector-dependent
tension by an average value that applies for all wave vectors. This
certainly changes any quantitative predictions, but does not influence
the qualitative aspects of those results.  This naive estimate should
evidently be replaced by a much more sophisticated dynamic analysis
that simultaneously performs the dynamic coarse-graining in the
presence of the advective non-linearity and a self-consistent (or
coarse-graining) procedure to recover the Helfrich interaction
behavior that stabilizes the lamellar stack. Such a calculation is
beyond the scope of this work.

If the layer spacing \textit{does} adjust in flow due to an induced
tension, a non-Newtonian response is likely to be found. Most probably
this will be shear thinning, because of the greater local regularity
of the flow, although it is not obvious that this is the case. The
magnitude of the viscous response is a complicated balance of
dissipation incurred within bilayers, and local inhomogeneous shears
due to the fluctuating layers. The single study that reported a change
in layer spacing also reported a shear thinning response
\cite{YamaTana95b}.  Shear thinning behaviour has been observed in
some Helfrich-stabilized systems including $C_{12} E_5$
\cite{YamaTana95b}, AOT \cite{asophie} and SDS
\cite{lerougeunpub}. 
\subsection{Effective Long Wavelength Dynamics }
The effective long wavelength dynamics of the single membrane is of
the form (in Fourier space)
\begin{widetext}
\begin{subequations}
\begin{gather}
\label{eq:90}
  -i\omega h^<(\bq, \om ) +i \gd \sum_{\Omega} 
  \sum_{\bk}(q_x -k_x ) h^<(\bk , \Om) h^<(\bq - \bk , \om - \Om) 
   = -\left[\Lambda(\bq)\kappa q^4\, + \Lambda_x q_x^2\right]
  h^<(\bq) +\xi^< (\bq, \om ), \\
\langle \xi^< \left(\bq_1, \om_1 \right) \xi^< \left(\bq_2,
  \om_2\right)\rangle =(D(\bm{q}_1) + 
  D_xq_x^2)\delta(\bm{q}_1 + \bm{q}_2)\delta(\omega_1 + \omega_2),
\end{gather}
\end{subequations}
where $h^<(\bq, \om )$ is the small wave vector (coarse-grained) height
field, and the noise $\xi^<$, in principle, incorporates the eliminated
degrees of freedom in addition to the original small scale degrees of
freedom. This yields a proportionality between correlation and
response, and a generalized fluctuation-dissipation theorem (FDT) is
satisfied, although the simple proportionality factor of temperature
relating correlation and response is replaced by the more complicated
noise correlations. In the case, which we have assumed, that the
scaling limit is reached before the patch size has been reached, $D_x$
vanishes and an effective temperature, albeit shear rate dependent,
can be ascribed to the system according to the fixed point value for
$D$. 

Ideally, coarse-graining should continue until all wavelengths less
than the collision length $L_p$ have been removed, at which point the
resulting theory would be used as a starting point for understanding
the dynamics of the usual mean smectic layer displacement $u$, rather
than the microscopic membrane position $h$.  Note that, at this point,
collisions intervene in a non-trivial way to limit affine layer
advection, and the coarse-grained smectic phase variable $u$ advects
according to $\dot{\gamma} y\partial_x\,u$ rather than $\dot{\gamma}
u\partial_x u$.  This behavior should, in principle, emerge smoothly in
an ideal calculation.

The resulting dynamics of a strongly fluctuating layered system
in shear flow are best cast in terms of the velocity field, in the
standard two-fluid form \cite{RamaswamyPCL93}, as 
\begin{align}
   \rho\left(\partial_t + \bm{v}\cdot\bm{\nabla}\right)\bm{v} &= 
    -\bm{\nabla} p + \eta\nabla^2\bm{v} + \hat{\bm{n}}\,f_n\\
\left(\partial_t + \bm{v}\cdot\bm{\nabla}\right)u &= v_z,
\end{align}
where we have, for convenience, shown the form in the absence of
permeation. The normal force $f_n$ differs from the usual normal force
by the term generated upon coarse-graining,
\begin{equation}
  \label{eq:91}
  f_n = -\left[\frac{\delta \cal{F}}{\delta u(q,t)} +
\frac{\Lambda_x\Lambda^{-1}(\bq)}{d} q_x^2 u(q,t)\right] ,
\end{equation}
\end{widetext}
where the free energy $\cal{F}$ should also include the layer compression
energy density $\tfrac12 \bar{B}\left(\partial_z u\right)^2$. Note the
factor of $d$ in the second term, reflecting the inherent three
dimensional nature of smectic elasticity.  The noise defines an
effective temperature that is generally not the physical temperature,
and may have additional correlations $D_x q_x^2$ that reflect the flow
(depending on whether or not the scaling regime has been reached).

The additional term is only present for strong flows, and penalises
layer undulations in the $x$ direction; this is because such
undulations are performed at the expense of the microscopic height
fluctuations, which are highly stretched in strong flows. This term is
\textit{not} expected to appear in situations where the microstructure
of the smectic layers is essentially undisturbed by flow, as in
typical thermotropic smectics (but see the calculation of Auernhammer
\textit{et al.}  \cite{AuernhammerBP00} for a counter example). One
could also envision this term as a non-equilibrium contribution to an
effective free energy, which has been postulated by Jou and co-workers
in their studies of complex fluids using extended irreversible
thermodynamics \cite{Criado-Sancho.Jou.ea98}; however the dependence
on strain rate that we derive,
$\Lambda_x\sim\dot{\gamma}^{\varepsilon_m}$, is not necessarily
analytic, unlike their assumptions.

It is important to remember that the generation of the
tension-restoring term is only one of several possible dynamic
effects; other effects include the rearrangement of defect
distributions, which is also likely to lead to a shear thinning
response \cite{MeyerABK00}.

\subsection{Summary}
\label{sec:summary}
In this work we have studied the effect of flow on the dynamics
of fluctuating membranes. We have made several assumptions, which
we collect for completeness:
\begin{itemize}
\item We assumed that an $\varepsilon$-expansion is sufficient to
  describe the effect of coarse-graining the theory up to the
  collision length; in this limit the renormalized noise reduces to an
  effective temperature.  Whether or not scaling is truly reached is
  an open question. It is more likely that there are residual noise
  correlations when the collision length has been reached. Moreover,
  the critical dimension $d_c$ is quite high and fluctuations are
  quite important; we have considered $m=-1,0,2$, for which,
  respectively, $d_c=\tfrac{15}{2},8,9$.
\item Since an $\varepsilon$-expansion is not likely to hold so far
  from the critical dimension, our calculation is strictly a
  self-consistent one-loop calculation.
\item We have considered the different membrane relaxation mechanisms
  (permeable, squeezing, isolated) separately. In reality, the
  mechanism changes during the coarse-graining process, according to
  Fig.~\ref{fig:0a}; nonetheless, this does not detract from our
  primary message, that the perturbation of the microstructure of
  highly fluctuating membranes can lead to an additional restoring
  term in the long wavelength dynamics.
\item The coarse-graining can be performed only up to $L_p$, because
  at this length scale the long range steric repulsion is important.
  In fact, we have completely ignored steric interactions.  A more
  precise treatment would involve simultaneously treating flow and
  collisions, or treating the flow within a self-consistent scheme
  using, for example, a harmonic potential to mimic collisions.
\end{itemize}

The significant accomplishments of this study have been, first, a
qualitative estimate of the effect of flow on highly fluctuating
lamellar phases. More importantly, however, is the demonstration that
flow can strongly modify the fluctuation spectrum and generate new
effects in the macroscopic response, via an RG-like self-consistent
coarse-graining technique. The theory that emerges has a natural
effective noise that need not satisfy the usual equilibrium
fluctuation-dissipation theorem. Similar renormalizations of
hydrodynamic descriptions can be expected for other complex fluid
systems with highly fluctuating mesoscopic degrees of freedom, such as
wormlike micellar systems (for example, the micellar length, modulus,
and relaxation times could be expected to renormalize due to the
effect of flow on the local charge distribution and undulation
spectra).

\begin{acknowledgements}
  We thank A.~J. Bray for helpful advice.
\end{acknowledgements}
\appendix
\section{Permeation length scale} \label{sec:perm-length-scale}
\begin{figure*}[!htb]
  \includegraphics[width=10cm]{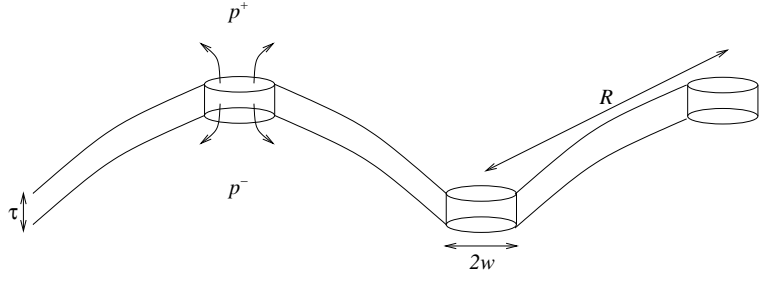}
     \caption{Pore defects in a membrane (side view).}
      \label{fig:10}
\end{figure*}
We estimate the permeation length $\zeta$ by assuming that permeation
is dominated by solvent flow through pores in the membranes; indeed, at
certain surfactant concentrations pores are very common -- see for
example the studies \cite{ZipfelBLR99,BKL94}.
We assume cylindrical pores of diameter $2 w$ within a membrane of 
thickness $\tau$, separated
by a mean distance $R$ (Fig.~\ref{fig:10}).  We wish to derive the
kinetic coefficient $\Lambda_0$ for layer relaxation, defined by
\begin{equation}
  \label{eq:13}
  \partial_t h = -\Lambda_0 \frac{\delta \cal{H}}{\delta h}. 
\end{equation}
Identifying the mean solvent flow velocity $\langle v \rangle$ with the
relaxation speed of the membrane $\partial_t h$ yields
\begin{equation}
  \label{eq:70}
  \langle v \rangle \sim \Lambda_0 \frac{\delta \cal{H}}{\delta h}.
\end{equation}
The pressure difference $\Delta p=p^+-p^-$ driving solvent flow  
is given by the force per area $\delta {\cal H}/\delta h$ that 
the membrane exerts on the fluid. Thus the kinetic coefficient is
\begin{equation}
  \label{eq:71}
  \Lambda_0 \sim \frac{\langle v \rangle}{\Delta p}.
\end{equation}
For each pore the mean velocity is the flux of material $Q$ flowing
through a pore per unit surface area of membrane $A$,
\begin{equation}
  \label{eq:40}
  \langle v \rangle = \frac{Q}{A}.
\end{equation}
We note that in terms of permeability $\cal P$, the flux is given by
$Q/A=-{\cal P}\nabla p /\eta$ (Darcy's law).  Thus ${\cal P} \simeq
\Lambda_0 \eta \tau$, which was used by Leng \cite{LengNR01} in the
context of swelling compressed lamellar phases.  To calculate the flux
we assume that the pressure gradient sets up a Poiseuille flow
given by the viscous flow force balance
\begin{equation}
  \label{eq:72}
  \eta\frac{v_c}{w^2} \simeq  \frac{\Delta p}{\tau} 
\end{equation}
where $v_c$ is the velocity of the solvent at the centre of the flow,
and so
\begin{equation}
  \label{eq:73}
 \frac{Q}{A} \sim \frac{w^4}{R^2 \tau}\frac{\Delta p}{\eta}.  
\end{equation}
On substituting the expression for the mean velocity in
Eq.~(\ref{eq:71}) by the flux in Eq.~(\ref{eq:73}) we obtain an
expression for the kinetic coefficient
\begin{equation}
  \label{eq:75}
   \Lambda_0\sim \frac{w^4}{\tau R^2 \eta}.
\end{equation}
Therefore, the ``permeation length'' is $\zeta=w^4/(R^2 \tau)$.
\section{Calculation of the Self-Energy and the Critical Points 
  for different relaxation mechanisms}
\label{sec:self-energy}
\begin{widetext}
In this appendix we calculate the coarse-grained self-energy
$\Sigma(\bq, 0)$, which we insert into the recursion relation
(Eq.~\ref{eq:27}) to find the 
fixed points $\Lambda_x^*$ for different values of $m$.  We also calculate the
general expression for the self-energy, and the associated $\Lambda_x$
derived from the recursion relation in a ``first step'' 
coarse-graining process.

\subsection{Coarse-grained self energy}
After demonstrating that a naive calculation of the self-energy,
equivalent to $\Lambda_x=0$, leads to a divergence, the main task here
is to show how the divergence is eliminated for $\Lambda_x\neq 0$,
enabling us to proceed with the integration.  Having established in
Section \ref{sec:one-loop-correction} that the self energy
(Eq.~\ref{eq:13d}) can be written in powers of $q_x^2$, we commence
from its expression in terms of the renormalized propagator
(Eq.~\ref{eq:63}) and its corresponding noise correlation
(Eq.~\ref{eq:62});
\begin{equation}
  \label{eq:75b}
  \Sigma(\bq , \om )= -\gd^2 \sum_{\Om } \sum_{\bk }
  q_x \left(\frac{q_x}{2}-k_x \right) G_R\left(\frac{\bq}{2}-\bk ,\Om \right) 
   \left|G_R \left(\frac{\bq}{2}+\bk ,\om-\Om \right) \right|^2 
   \left[D_0\left|\frac{\bq}{2}+\bk\right|^m 
   + D_x\left(\frac{q_x}{2}+k_x\right)^2\right].
\end{equation}
We consider the slow hydrodynamic regime $\omega\to 0$, and ignore
frequency-dependent corrections, so that
\begin{equation}
  \label{eq:54}
  G_R(\bq , \om )^{-1}\simeq -i\om + \Lambda(\bq)  S^{-1}(\bq)+\Lambda_x q_x^2,
\end{equation}
where $\Lambda_x\equiv a_2(\bq, 0)$, $\Lambda(\bq)=\Lambda_0 q^m$ and
$S^{-1}(\bq)=\kappa q^4$. The disregard for frequency-dependent
corrections should suffice to obtain scaling properties.  The sums are
converted to integrals by Eq.~(\ref{eq:3})
\begin{equation}
  \label{eq:33}
  \sum_{\Omega}\sum_{\bk}\to 
  \int\frac{d\Omega}{2\pi} 
  \int_{\pi/L_p}^{\pi/a}\frac{d^{d-1}k S_{d-1}}{(2\pi)^{d-1}},
\end{equation}
where $S_{n}=n\pi^{n/2}/\Gamma(n/2+1)$ is the surface area of an
$n$ dimensional unit sphere and the limits in $\bk$ are given by the
physical cutoffs. The first step is to perform the 
$\Omega$-integral by contour integration, yielding
a positive pole at $i\Omega=-\left[\Lambda_x\left(k_x+\frac{q_x}{2}\right)^2
+\kappa\Lambda_0\left|\bk+\frac{\bq}{2}\right|^{4+m}\right]$, 
\begin{equation}
  \label{eq:9}
  \Sigma (\bq ,0) = -\frac{\gd^2}{2}\!\!
  \int_0^{\infty}\!\!\frac{k^{d-2}dk S_{d-1}}{(2\pi)^{d-1}}
  \frac{q_x\left(\frac{q_x}{2}-k_x\right)
  \left[D_0\left|\bk+\frac{\bq}{2}\right|^m
  +D_x\left(k_x+\frac{q_x}{2}\right)^2\right]}
  {\left[\Lambda_x\left(k_x+\frac{q_x}{2}\right)^2
  +\kappa\Lambda_0\left|\bk+\frac{\bq}{2}\right|^{4+m}\right]
  \left[2\Lambda_x\left(k_x^2+\frac{q_x^2}{4}\right)
  +\kappa\Lambda_0\left|\bk+\frac{\bq}{2}\right|^{4+m}\!\!
  +\kappa\Lambda_0\left|\bk-\frac{\bq}{2}\right|^{4+m}\right]}.
\end{equation}
As we explained in Section~\ref{sec:one-loop-correction}, the leading
long wavelength behaviour arises from an expansion of Eq.~(\ref{eq:9})
to lowest order in $q_x^2$, which we now investigate for the cases
$\Lambda_x=0$ and $\Lambda_x\neq0$. For $\Lambda_x=0$, and thus also
$D_x=0$, the self-energy becomes
\begin{equation}
  \label{eq:75c}
  \Sigma(\bq, 0)= 
  -\gd^2 q_x^2\frac{D_0}{2(\kappa\Lambda_0)^2}  
  \int_0^{\infty}\frac{k^{d-2} dk S_{d-1}}{(2\pi)^{d-1}} 
  \left[\frac{1}{4|\bk|^{8+m}}+\frac{2k_x(k_x^3+k_x
      k_{\perp}^2)}{2|\bk|^{12+m}}\right] + {\cal O\/}(q_x^4),
\end{equation} 
which diverges at low $\bk$ for $10+m-d>0$ (note that all 
relaxation mechanisms we consider obey $m>d-10$). As we discuss at the end of
Section~\ref{sec:one-loop-correction}, this divergence is
unphysical. Upon coarse-graining the theory, the term $\Lambda_x
q_x^2$ will be generated, which obviously changes the character of the
integral.
Next we show that the implementation of coarse-graining such 
that $\Lambda_x\neq0$ removes this divergence.

Following the first step of the renormalization group analysis in 
Section~\ref{sec:renorm-group-rg} the removal of high wave vectors 
$\bk^>$ in the $x$
direction in the range $\lambda e^{-l}<k_x<\lambda$ is equivalent to 
a change in limits, 
\begin{equation}
  \label{eq:80b}\sum_{\bk^>} \to 
  \int_{\lambda e^{-l}}^\lambda\frac{dk_x}{2\pi}
  \int_0^{\infty}\frac{d^{d-2}k_{\perp} S_{d-2}}{(2\pi)^{d-2}}.
\end{equation}
Note that there is no change in the limits in the perpendicular
direction.  We showed in Section~\ref{sec:renorm-group-rg} that when
$\Lambda_x\neq0$ and $m\geq-1$, we approximate $|\bk|\to|\bk_{\perp}|$
so that the renormalized propagator becomes Eq.~(\ref{eq:16}) and the
noise becomes Eq.~(\ref{eq:17}).  In addition, within the scaling
regime or close to the fixed point, $D_x\to0$.  With these assumptions
the resulting approximation to the lowest order expansion of the self
energy is
\begin{align}
  \Sigma (\bq ,0) & = -\gd^2 q_x^2 \frac{D_0}{2} 
  \int_{\lambda e^{-l}}^{\lambda} \frac{dk_x}{2\pi} 
  \int_0^{\infty}\frac{k_{\perp}^{d-3}dk_{\perp}S_{d-2}}{(2\pi)^{d-2}}.
  \frac{k_{\perp}^m}{\left(\Lambda_x k_x^2 + \kappa \Lambda_0
    k_{\perp}^{4+m}\right)^2} \left[\frac14 + \frac{\Lambda_x k_x^2}
    {2\left(\Lambda_x k_x^2 +\kappa \Lambda_0
        k_{\perp}^{4+m}\right)}\right] + {\cal O\/}(q_x^4).\label{eq:48}
\end{align}
The divergence that we encountered before for $\Lambda_x=0$ is now eliminated,
which enables us to proceed with the integration over all wave vectors
in the transverse direction, in order to
calculate the self-energy $\Sigma (\bq ,0)$ to be inserted into the
recursion relation (Eq.~\ref{eq:27}). 

\subsection{Calculation of the fixed points}

With the condition that $d+m-2> 0$ (a criterion that we discuss later),
integration of Eq.~(\ref{eq:48}) leads to
\begin{equation}
 \label{eq:31b}
 \Sigma (\bq ,0)= 
 -2\frac{(14+2m-d)}{m +10 -d}U\Lambda_x q_x^2 l  \qquad 
 \qquad \left(d+m-2 > 0\right),
\end{equation}
where
\begin{equation}
  \label{eq:32}
  U = \gd^2 \frac
  {S_{d-2}}{8(m+4)(2\pi)^{d-2}}\Gamma\left(\frac{d+m-2}{m+4}\right) 
  \Gamma\left(\frac{14+2m-d}{m+4}\right)D_0 
  (\kappa\Lambda_0)^{\frac{2-d-m}{m+4}}\Lambda_x^{\frac{-14-2m+d}{m+4}}
  \lambda^{\frac{-16-m+2d}{m+4}}.
\end{equation}
$\Gamma(v)$ is the Gamma function and we have used
\begin{equation}
  \label{eq:34}
  \int_0^{\infty}\frac{x^{z-1}dx}{(1+x)^{z+w}}=\frac{\Gamma(z)\Gamma(w)}
  {\Gamma(z+w)}, \qquad (\Re(z) >0, \Re(w)>0).
\end{equation}
The negative exponent of $\Lambda_x$ in Eq.~(\ref{eq:32})
reveals the divergence established earlier and therefore 
reaffirms the necessity for coarse-graining.

Inserting the self energy (Eq.~\ref{eq:31b}) into Eq.~(\ref{eq:27})
leads to an expression for the recursion relation for $\Lambda_x$,
which is most conveniently written in terms of the coupling constant
$U$:
\begin{equation}
  \label{eq:35}
  \frac{dU}{dl}=\frac{16+m-2d}{m+4}U 
  -2\frac{(14+2m-d)^2}{(m+4)(m+10-d)}U^2,
\end{equation}
where we have used Eq.~(\ref{eq:25}) to eliminate $z$.  Having thus
removed the high wave vectors and rescaled all the parameters the final
step of the RG analysis is to find the fixed points of
Eq.~(\ref{eq:35}) for which the theory is invariant.  Consistent with
the previous determination of the critical dimension, the linear term
changes sign for $d=d_c=(16+m)/2$. Since the quadratic term is
negative for $d<m+10$ (or $d<d_c$ for $m<-4$), there is a non-zero
stable fixed point
\begin{equation}
  \label{eq:36}
  U^*=\frac{\epsilon}{9(d_c -6)} + \ldots
\end{equation}
to first order in $\epsilon$ where $\epsilon=d-d_c$. 
This shows that the RG perturbation
is well-behaved and the exponents are correct.  For $d>d_c$ the only
stable fixed point is $U^*=0$, corresponding to an irrelevant
non-linearity and recovering the exponents for the linear theory.
Thus in the general case for $d+m-2>0$, 
\begin{align}
 \label{eq:36a}
 \Lambda^*_x &=
 \alpha_{m+1}\left[T^{4+m}\kappa^{2-d-m}\lambda^{-16-m+2d}
             \Lambda_0^{-2m-8}\gd^{2(4+m)}\right]^{1/(14+2m-d)}, \\
 \text{where} \qquad \alpha_{m+1} &= 
 \left[\frac{9 S_{d-2}}{4(m+4)(2\pi)^{d-1}}\left(\frac{4+m}{16+m-2d}\right)
 \Gamma\left(\frac{d+m-2}{m+4}\right)
 \Gamma\left(\frac{14+2m-d}{4+m}\right)\right]^{(4+m)/(14+2m-d)} .
\end{align} 
We may apply this result to two of the relaxation mechanisms that we
considered; both results are found in Table~\ref{tab:1} in Section~\ref{sec:disc-flows-coarse}. For the permeable case, $d=3$ and $m=0$, 
\begin{align}
  \label{eq:36b}
  \Lambda^*_x &= \alpha_{1}
  \frac{T^{4/11}\Lambda_0^{3/11}}{\kappa^{1/11}\lambda^{10/11}} \gd^{8/11},   
  &\text{where} \qquad \alpha_{1} &=\left[\frac{189 S_1}{2560\pi^2}
    \Gamma\left(\frac14 \right)\Gamma\left(\frac34
    \right)\right]^{4/11}\simeq 0.342\,\,.  \\
  \intertext{For the confined case, $d=3$ and $m=2$,}
  \Lambda^*_x&= \alpha_3 
  \frac{T^{2/5}\Lambda_0^{4/5}}{\kappa^{1/5}\lambda^{4/5}} \gd^{4/5},
  &\text{where} \qquad \alpha_3 &=\left[\frac{45 S_1}{256\pi^2}
    \Gamma^2\left(\frac12 \right)\right]^{2/5} \simeq 0.378\,\,. \label{eq:36c}
\end{align}

A similar analysis cannot be conducted for the case $d=3, m=-1$.  In
the hydrodynamic limit a divergence in the lower limit in the
$k_{\perp}$ integral of Eq.~(\ref{eq:48}) occurs for $d+m-2\leq 0$.
By introducing a lower cutoff given by the inverse collision length of
the system $L_p^{-1}$ and writing the integral in terms of the
dimensionless quantity $y=L_p^3 \Lambda_x \lambda^2/\kappa\Lambda_0$,
the expression for the self energy for $d=3, m=-1$ becomes
\begin{equation}
  \label{eq:5}
   \Sigma (\bq ,0) = -\frac{\alpha_0\gd^2 D_0}{\lambda^3\Lambda_x^3}
   \left[3\left(\ln(y+1)-\frac{y}{y+1}\right)-\frac{y^2}{(y+1)^2}\right]
   \Lambda_xq_x^2 l , \qquad \text{where} \qquad 
   \alpha_0=\frac{S_1}{96\pi^2}\simeq 0.000528. 
\end{equation}
We now proceed as before by combining Eq.~(\ref{eq:27}) and the 
expression for the self-energy
(Eq.~\ref{eq:5}) to give the differential flow equation in terms of $y$,
\begin{equation}
  \label{eq:4}
  \frac{dy}{dl}=y\left[-1+\frac{\alpha_0\gd^2 D_0}{\lambda^3}
  \left(\frac{L^3_p\lambda^2}{\kappa\Lambda_0 y}\right)^3 
 \left[3\left(\ln(y+1)-\frac{y}{y+1}\right)-\frac{y^2}{(y+1)^2}\right]\right]. 
\end{equation}
The unstable fixed point corresponding to the irrelevant 
non-linearity is given by $\Lambda_x=0$.  As we are unable to give an analytic
expression for the stable fixed point solution of 
Eq.~(\ref{eq:4}) we show instead its power series in $\gd^2$,
\begin{equation}
  \label{eq:53}
  \Lambda^*_x =\alpha_0 \frac{T L_p^6 \lambda}{\kappa^2
    \Lambda_0}\gd^2
  \left[1-\frac32\left(\alpha\frac{L_p^9\lambda^3}{\kappa^3\Lambda_0^2}\right)^2\gd^4 + {\cal O\/}(\gd^5)\right] . 
\end{equation}
Hence for small strain rates $\Lambda_x^*\sim\gd^2$.
 
\subsection{First step coarse-graining}

Here we demonstrate the procedure for a ``first step'' coarse-graining
of the self-energy in Section~\ref{sec:first-step}, i.e. we calculate
$\Lambda_x$ by perturbing about the trivial fixed point
$\Lambda_x^*=0$.  First we return to the expression for the
coarse-grained self-energy in Eq.~(\ref{eq:9}) derived from
Eq.~(\ref{eq:75b}). Upon coarse-graining in the $x$ direction, the
limits of the sum and the integration are changed according to
Eq.~(\ref{eq:80b}). However, due to being far from the scaling regime,
the assumption that $|\bk|\to|\bk_{\perp}|$ no longer holds. Hence in
the limit $\Lambda_x\to0$ (and thus $D_x\to0$) appropriate for a
``first step'' coarse-grain, the integral becomes
\begin{equation}
  \label{eq:38}
  \Sigma(\bq, 0)= 
  -\gd^2 q_x^2\frac{D_0}{2(\kappa\Lambda_0)^2}  
  \int_{\lambda e^{-l}}^{\lambda} \frac{dk_x}{2\pi} 
  \int_0^{\infty}\frac{k^{d-3} dk S_{d-2}}{(2\pi)^{d-2}} 
  \left[\frac{1}{4|\bk|^{8+m}}+\frac{2k_x(k_x^3+k_x
      k_{\perp}^2)}{2|\bk|^{12+m}}\right] + {\cal O\/}(q_x^4)
\end{equation}
from which we may compute the self-energy for general $m$ and $d$,
\begin{equation}
  \label{eq:55}
  \Sigma(\bq ,0) = -\beta\frac{T}{\kappa^2 \Lambda_0
    \lambda^{9+m-d}}\gd^2 q_x^2 l .
\end{equation}
$\beta$ is a numerical prefactor that depends on the relaxation
mechanism $m$ and the dimension $d$,
\begin{equation}
 \begin{split} 
  \label{eq:57}
  \beta = \frac{S_{d-2}}{8 (2\pi)^{d-1}}& \Gamma\left(\frac{d-2}{2}\right)
  \Gamma\left(\frac{10+m-d}{2}\right)\left[\Gamma\left(\frac{12+m}{2}\right)\right]^{-1} \\
  & \times
  \left[5\left(\frac{12+m-d}{2}\right)\left(\frac{10+m-d}{2}\right)
    +6\left(\frac{d-2}{2}\right)\left(\frac{10+m-d}{2}\right)+\frac{d}{2}
    \left(\frac{d-2}{2}\right)\right] .
 \end{split}
\end{equation}
On substituting Eq.~(\ref{eq:55}) into the recursion relation for 
around the trivial fixed point (Eq.~\ref{eq:28}) we find that,
\begin{equation}
  \label{eq:30}
   d\Lambda_x(l)\simeq\beta\frac{T}{\kappa^2 \Lambda_0
     \lambda^{9+m-d}}\gd^2 l, 
\end{equation} 
which leads to the ``tension'' that depends on the cutoff in
Eq.~(\ref{eq:56}).
\end{widetext}

\end{document}